\documentclass[12pt,titlepage,a4paper]{article}
\usepackage{fancyheadings}
\usepackage{epsfig}
\usepackage{amssymb}
\usepackage{amsmath}
\usepackage{amsfonts}
\usepackage{color,graphicx}
\newtheorem{thm1}{Theorem}

\title {\sc {\bf The B-Exponential Map: \\ A Generalization of the Logistic Map, \\ and Its Applications In Generating \\ Pseudo-random Numbers}}
\author{Mahesh C Shastry \\ Dept. of Electrical and Electronics Engineering \\ National Institute of Technology Karnataka Surathkal \\ {\bf mahesh.shastry@gmail.com} \\ \\  Nithin Nagaraj \\ School of Natural and Engineering Sciences \\ National Institute of Advanced Studies\\ {\bf nithin\_nagaraj@yahoo.com} \\ \\  Prabhakar G Vaidya \\ School of Natural and Engineering Sciences \\ National Institute of Advanced Studies\\ {\bf pgvaidya@nias.iisc.ernet.in}}
\date {July 13, 2006}
\begin{document}
\maketitle
\begin{abstract}
A 1-dimensional generalization of the well known Logistic Map is
proposed. The proposed family of maps is referred to as the
B-Exponential Map. The dynamics of this map are analyzed and found
to have interesting properties. In particular, the B-Exponential
Map exhibits {\it robust chaos} for all real values of the
parameter $B \geq e^{-4}$.
\par We then propose a pseudo-random number generator based on the B-Exponential
Map by chaotically hopping between different trajectories for
different values of B. We call this BEACH (B-Exponential
All-Chaotic Map Hopping) pseudo-random number generator. BEACH
successfully passes stringent statistical randomness tests such as
ENT, NIST and Diehard. An implementation of BEACH is also
outlined.
\end{abstract}

\section{Introduction}
A function which has the same domain space and range space will be
called a map. A chaotic map $x_{n+1}=F(x_n)$ is typically a
non-linear discrete dynamical iteration equation, which exhibits
some sort of chaotic behavior. Chaos is characterized by
deterministic nonlinearity, non-periodicity, sensitive dependence
on initial conditions, boundedness and topological
transitivity~\cite{yorke}. One of the well known 1-dimensional
iterative maps which exhibits chaotic properties is the Logistic
Map~\cite{yorke}. The Logistic Family is defined by the iteration
$x_{n+1}=ax_n(1-x_n)$. The parameter $a$ controls the dynamics of
this map. The map attains full chaos at $a=4$ since it becomes
surjective at this value. Henceforth, we shall refer to the map
$x_{n+1}=4x_n(1-x_n)$ as the Logistic Map. The Logistic Map was
first proposed as a demographical model and was later applied to
study population dynamics of species considering the twin effects
of reproduction and starvation. The parameter $a$ is a positive
number and represents a combined rate for reproduction and
starvation.

\par In this report, we propose a generalization of the Logistic Map.
The generalized family of maps is referred to as the B-Exponential
Map. Characteristics of the B-Exponential Map such as return maps,
bifurcation diagram, Lyapunov exponents, Schwarzian derivatives
are investigated. The investigations reveal interesting
properties, in particular we show that the B-Exponential Map
exhibits {\it robust chaos} for a large range of B.

\par One of the interesting features of Chaotic systems from an application point of view is that
they statistically mimic random white noise while remaining
deterministic. This property along with others (such as
topological transtivity and Ergodicity, robust chaos) provide the
necessary requirements for Chaotic systems to design strong
pseudo-random number generators (PRNG). The B-Exponential Map,
exhibits these properties (as we shall soon show) and hence can be
utilized to build a PRNG. An implementation of the B-Exponential
Map as a PRNG is outlined.

\section{The B-Exponential Map}
\begin{figure}[!hbp]
\centering
\includegraphics[scale=.6]{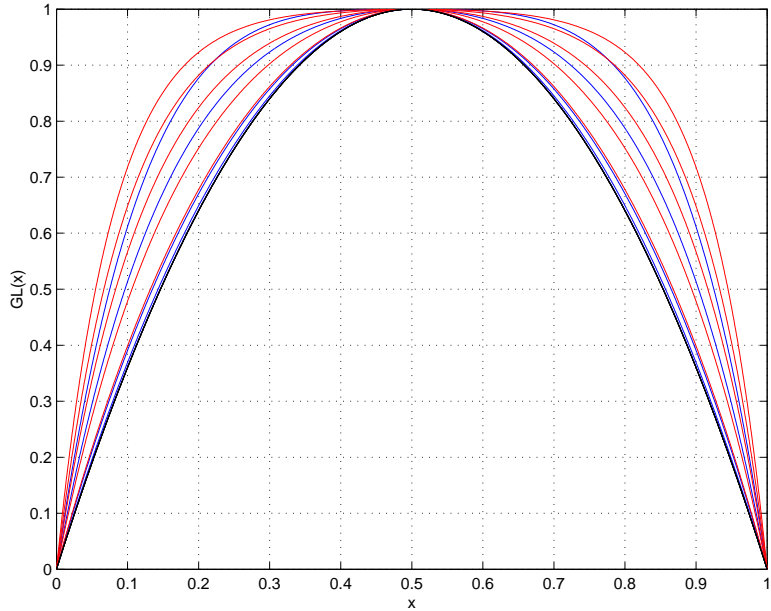}
\caption{The B-exponential map for various values of $B$. The blue
curves are are the maps for $B<1$, and the red ones for $B>1$, the
black curve is the logistic map. All the curves are for $B \geq
e^{-4}$.} \label{fig:many maps}
\end{figure}

The B-Exponential Map $GL(B,x)$ is defined as follows.

\begin{equation}
GL(B,x)=\frac{B-xB^x-(1-x)B^{1-x} } {B-\sqrt{B}} \label{eq:gl}
\qquad 0\leq x \leq 1 \textrm{~and~} B \in \mathbb{R}^+.
\end{equation}
\\Here, $B$ is the adjustable parameter. Note that
$x_{n+1}=GL(B,x_n)$ is the iteration function. A plot of $GL(B,x)$
is shown in Figures \ref{fig:many maps} and \ref{fig:many
maps_nonchaos} for different values of $B$. $GL(B,x)$ is unimodal
for $e^{-4} \leq B < \infty$.

Consider the interval $[0,1]$. $GL(B,x)$ is a linear combination
of $f(x) = xB^x$ and $f(1-x) = (1-x)B^{1-x}$ which are both
single-hump maps for all $B$. The critical point of $f(x)$ is
$\frac{-1}{\ln(B)}$ and that of $f(1-x)$ is $1 +
\frac{1}{\ln(B)}$. If $B>1$, these are negative and greater than
one respectively. Hence, we know that $GL(B,x)$ is unimodal for
$B>1$. For $B<1$, these two critical points lie within $[0,1]$ on
either sides of $x=0.5$. Consider $0<B<e^{-4}$. We show in
Appendix A that in this case,  we lose surjectivity, $x=0.5$ is a
local minimum and hence $GL(B,x)$ is no longer unimodal. For all
$B \ge e^{-4}$, we know that $x=0.5$ is a local maximum (at
$B=e^{-4}$, it is a point of inflexion). We know that $GL(B,0)=0$
and $GL(B,0.5) = 1$. Assume that there is a critical point in
$[0,0.5)$ and it is a local maximum. In such a case, there has to
be another critical point in $[0,0.5]$. This would mean that
$GL(B,x)$ will have at least 5 critical points in $[0,1]$ owing to
symmetry around $x=0.5$. Hence, $GL(B,x)$ is unimodal for $B \geq
e^{-4}$.

In this section, we explain some results concerning the
B-Exponential Map.

\begin{figure}[!hbp]
\centering
\includegraphics[scale=.6]{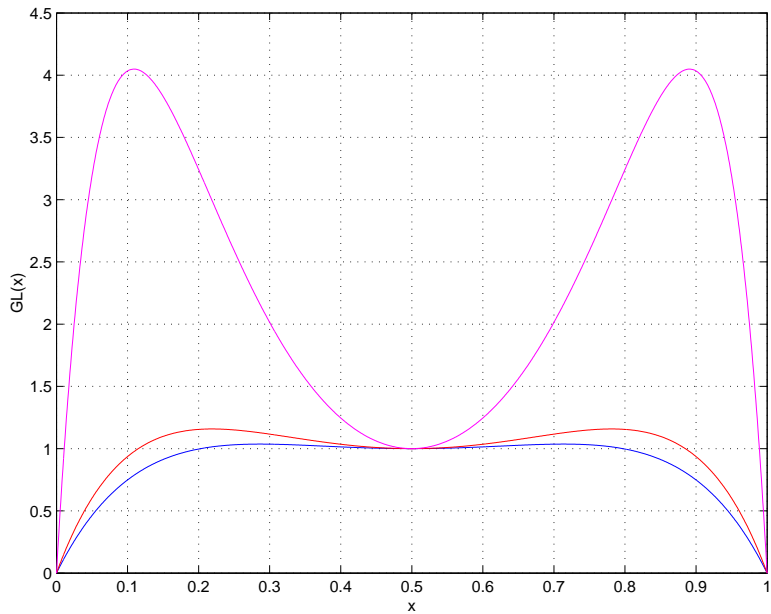}
\caption{The B-exponential map for various values of positive
$B<e^{-4}$.} \label{fig:many maps_nonchaos}
\end{figure}
%
%\begin{figure}[!hb]
%\centering
%\includegraphics[scale=.6]{first_ret_map.eps}
%\caption{The first return maps for various values of $B \geq
%e^{-4}$.} \label{fig:first return maps}
%\end{figure}
%
\newpage
The B-Exponential Map is a generalization of the Logistic Map
because of the following property:
\begin{equation}
\lim_{B \rightarrow 1} GL(B,x)=4x(1-x). \label{eq:g2}
\end{equation}
%%%
\\This interesting property can be derived by applying
L'Hospital's rule. The derivation is given in Appendix A.

The B-exponential map is concave for a wide range of $B$. We state
and prove the following properties.
\begin{thm1}
The B-Exponential Map is topologically conjugate to the Logistic
Map for $e^{-4} \leq B < \infty $. \label{thm:thm0}
\end{thm1}
\noindent {\bf Proof:~} There is only one critical point for
$GL(B,x)$ at $x=0.5$ for $e^{-4} \leq B < \infty $. We can define
a diffeomorphism of $GL(B,x)$ to the line $2x$ (standard Tent-map)
for $0 \leq x < 0.5$ by associating the appropriate heights.
Similarly, we can define another diffeomorphism to the line $2-2x$
for $0.5 \leq x \leq 1$. This establishes that $GL(B,x)$ is
topologically conjugate to the standard Tent-map which we know to
be topological conjugate to the Logistic Map. Thus, by transitive
relation of topological conjugacy, we know that $GL(B,x)$ is
topologically conjugate to the Logistic Map ($e^{-4} \leq B <
\infty $).
%
%
%\begin{thm1}
%The B-Exponential map is ergodic for all $B > e^{-4} $.
%\label{thm:thm1}
%\end{thm1}
%\noindent {\bf Proof:~} We know that the Logistic Map is ergodic
%for an invariant measure $\mu$. ??
%
%
\\
\par
\begin{figure}[!hbp]
\centering
\includegraphics[scale=.4,angle=270]{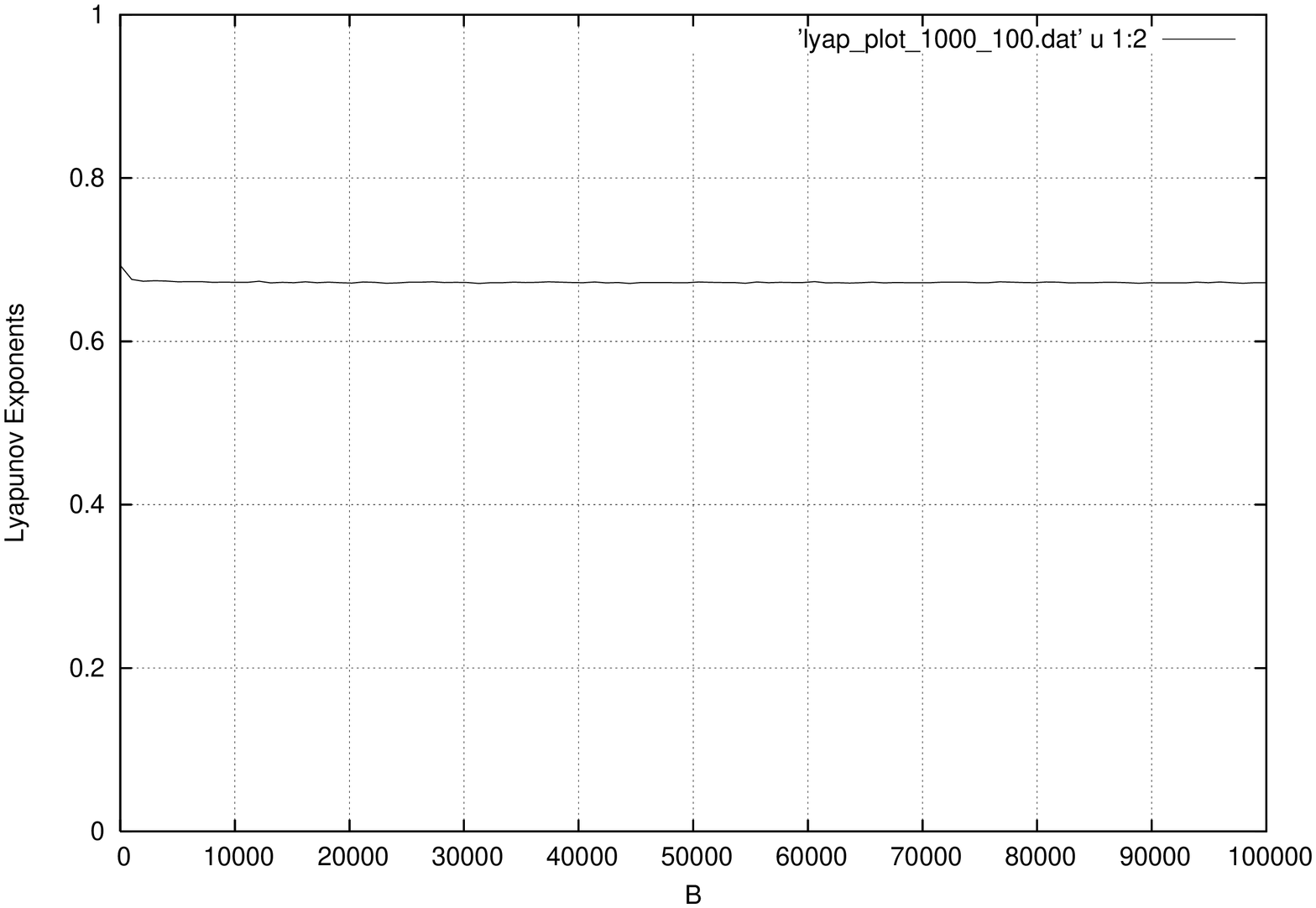}
\caption{A plot of the Lyapunov exponents of the B-exponential map
against the parameter B.}
 \label{fig:figlyap}
\end{figure}

 The Lyapunov exponents of the B-Exponential map appears to be
constant for all $B \geq e^{-4} $ and equal to $\ln{2}$. The
Lyapunov exponents are defined as follows~\cite{yorke}:
\begin{equation}
\lambda(B) = \lim_{T \rightarrow \infty} \frac{1}{T}
\sum_{t=0}^{T} \Big| \Big| \frac{d}{dx} GL(B,x) \Big|_{x=x_t}
\Big|.
\end{equation}
It is a difficult task to evaluate the above expression
analytically. We resort to numerical estimation of the Lyapunov
exponent. The plot of Lyapunov exponents of the B-Exponential Map
is shown in Figure \ref{fig:figlyap}. As it can been seen, the
Lyapunov exponent is ln2 for $B \geq e^{-4}$. The number of
iterations we used for computing the exponent is 10,000.
\begin{thm1}
The B-Exponential Map is chaotic for all real $B \geq e^{-4}$.
\label{thm:thm2}
\end{thm1}
\noindent {\bf Proof:~} There is no universal definition of Chaos.
We use the following conditions as necessary and sufficient for
Chaos:
\begin{itemize}
\item Determinism: $GL(B,x)$ is a deterministic map of $[0,1]
\rightarrow [0,1]$ for $B \geq e^{-4}$.%
 \item Surjective and Boundedness: $GL(B,x)$ is
bounded for all $0 \leq x \leq 1$. $GL(B,x)$ is surjective on
$[0,1]$ for $B \geq e^{-4}$ (see Appendix A). %
\item Sensitive dependence on initial conditions: $GL(B,x)$
exhibits sensitive dependence on initial conditions
(continuously). This is characterized by positive Lyapunov
exponents. $GL(B,x)$ seems to have a Lyapunov exponent of ln2=0.6931 (almost everywhere) for every $B \geq e^{-4}$ (refer to Figure~\ref{fig:figlyap}). %
\item Positive Topological Entropy: The symbolic dynamics of
$GL(B,x)$ is such that all possible transitions (0 to 0, 0 to 1, 1
to 0 and 1 to 1) are achieved. Here the Markov partitions are 0
(if $0 \leq x < 0.5$) and 1 (if $0.5 \leq x \leq 1$).  Hence
topological entropy is ln2
which is positive. %
\item Topological transitivity: Successive iterations of $GL(B,x)$
mixes the domain. For every pair of open sets $A,B \subseteq
[0,1]$, there is a $k>0$ such that $T^{(k)}(A) \cap B \neq \O$. %
\item Periodic points are dense in $[0,1]$: This follows by the
fact that  $GL(B,x)$ is topological conjugate to the Logistic Map.
\end{itemize}
\subsection{Robust Chaos}
Robust Chaos is defined by the absence of periodic windows and
coexisting attractors in some neighborhood of the parameter
space~\cite{Banerjee}. Barreto~\cite{Barreto} had conjectured that
robust chaos may not be possible in smooth unimodal
one-dimensional maps. This was shown to be false with
counter-examples by Andrecut~\cite{Andrecut} and
Banerjee~\cite{Banerjee}. Banerjee demonstrates the use of robust
chaos in a practical example in electrical engineering. Andrecut
provides a general procedure for generating robust chaos in smooth
unimodal maps.

\begin{figure}[!hbp]
\centering
\includegraphics[scale=.6]{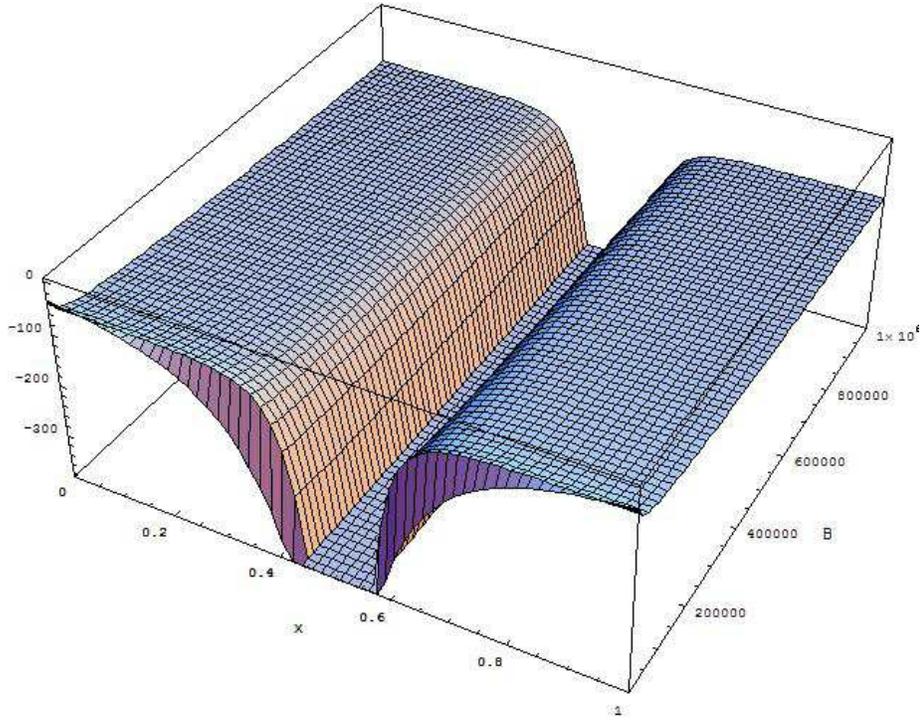}
\caption{Schwarzian derivative of $GL(B,x)$ plotted for $B=0$ to
$10^6$.} \label{fig:figschw3d}
\end{figure}

\begin{figure}[!hbp]
\centering
\includegraphics[scale=.6]{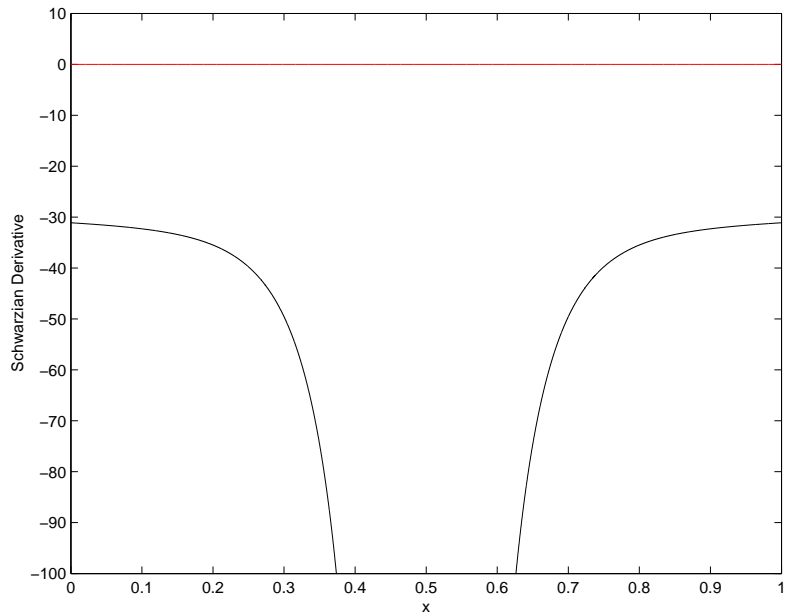}
\caption{Schwarzian derivative of $GL(B,x)$ for $B=1000$. It can
be observed that the value is always negative.}
\label{fig:figschw2d}
\end{figure}

As observed by Andrecut~\cite{Andrecut2}, robust chaos implies a
kind of ergodicity or good mixing properties of the map. This
makes it very beneficial for cryptographic purposes. The absence
of windows would mean that the these maps can be used in hardware
implementation as there would be no {\it fragility} of chaos with
noise induced variation of the parameters. We shall demonstrate
that the B-Exponential Map exhibits robust chaos and this property
makes it highly beneficial for generating pseudo-random number
generators.
\begin{thm1}
The B-Exponential Map exhibits {\it robust chaos}~ $B \geq
e^{-4}$. \label{thm:thm2}
\end{thm1}
\noindent {\bf Proof:~} We know that $GL(B\geq e^{-4},x)$ has only
one critical point at $x = 1/2$. We numerically find that the
Schwarzian derivative which is given by
\begin{equation}
S(f)(x)=\frac{f'''(x)}{f'(x)}-\frac{3}{2}\Big(\frac{f''(x)}{f'(x)}\Big)^2
\qquad whenever~f'(x)\neq 0.
\end{equation}

is negative (Figures~\ref{fig:figschw3d} and \ref{fig:figschw2d}).
Since $GL(B,x)$ is smooth and unimodal in the range $[e^{-4},
\infty]$, we invoke the theorem in \cite{yorke} to say that there
can be at most one attracting periodic orbit with the critical
point in its basin of attraction. Since $x=0.5$ is the only
critical point and it ends in the value 0 after two iterates, we
end up on an {\it unstable fixed point} (one can verify that
$|GL'(B,0)| > 1$). Hence, we infer that $GL(B,x)$ does not have
any attracting periodic orbits
for $B$ in $[e^{-4}, \infty]$.\\
It is interesting to observe that one can derive a generalization
of the standard Tent Map by noting that there is a conjugacy
between the Tent Map and Logistic Map. The Generalized Tent Map is
described in Appendix A. By means of topological conjugacy, one
could potentially generate a number of maps, all of which exhibit
{\it robust chaos}. These could also be used for designing PRNG,
but we shall restrict our attention to the B-Exponential Map in
this report.
\begin{figure}[!hbp]
\centering
\includegraphics[scale=.6]{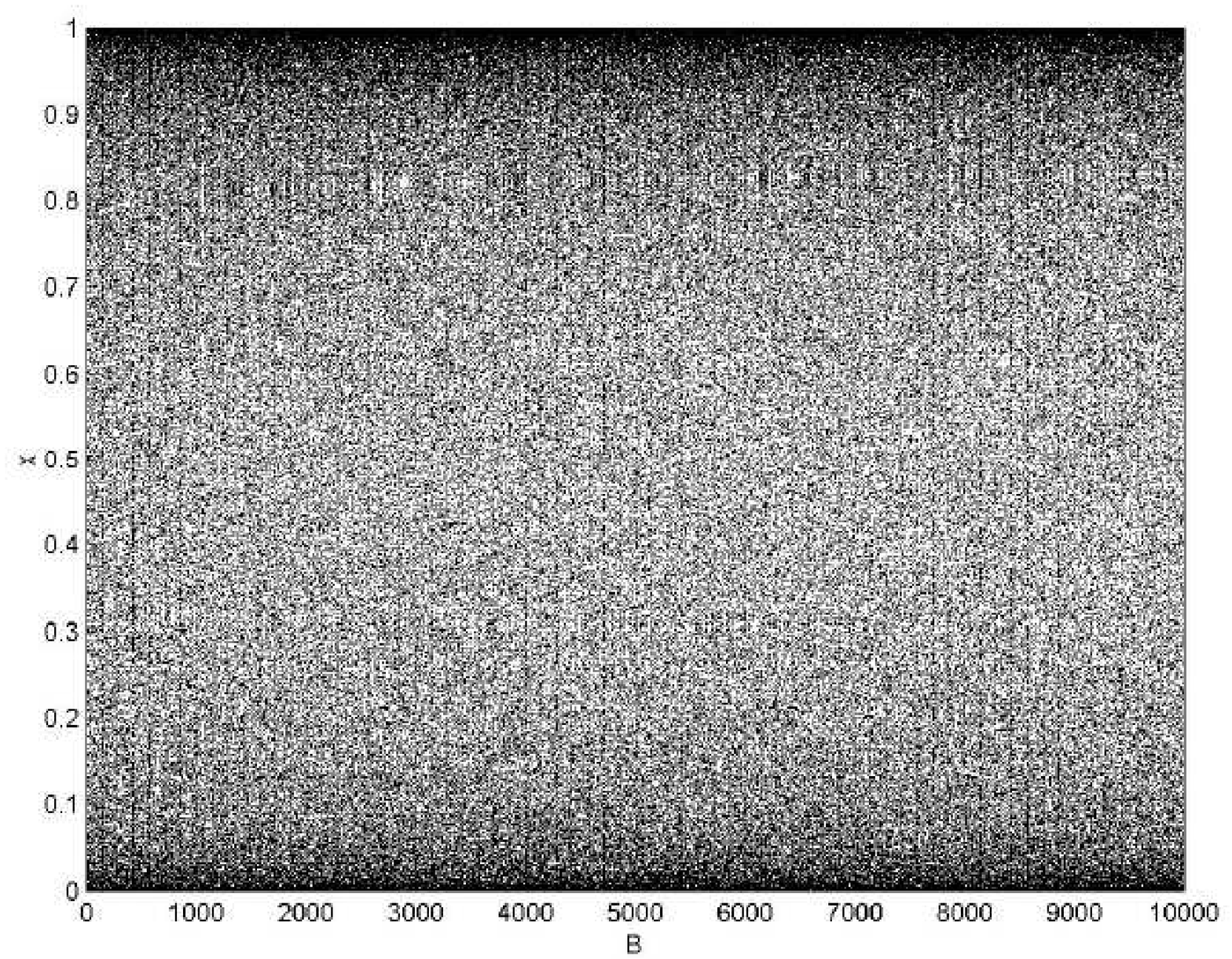}
\caption{The bifurcation diagram of the B-exponential map for B
ranging from near 0 to 10000. The transition to chaos at $e^{-4}$
is not visible because of the large range of B.}
\label{fig:figbif}
\end{figure}

\begin{figure}[!hbp] \centering
\includegraphics[scale=.6]{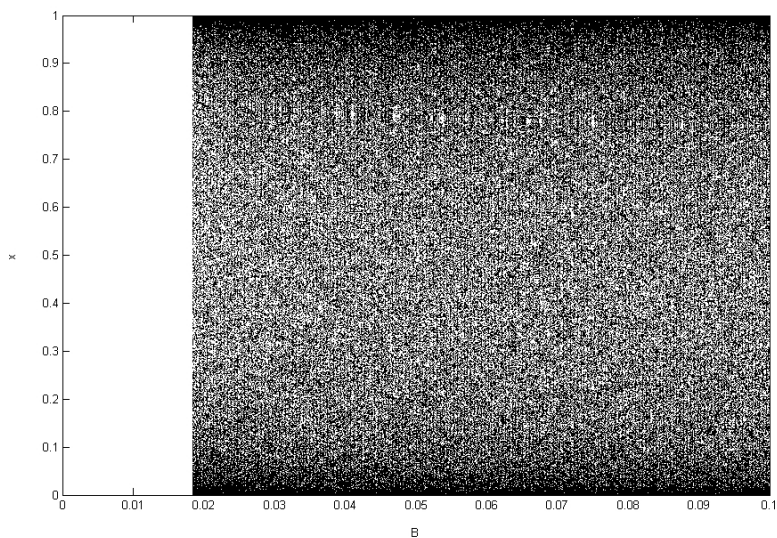}
\caption{The bifurcation diagram of the B-exponential map for B
ranging from close to 0 to 0.1. The breakdown of chaos is clearly
visible at $e^{-4}$, i.e about $0.018316$.} \label{fig:figbif
break}
\end{figure}

The bifurcation diagram of the B-Exponential Map is shown in
Figures \ref{fig:figbif} and \ref{fig:figbif break}. It is clear
from the bifurcation diagram that the map is chaotic for a large
range of B. This property is very interesting. This is a unique
continuous map in that respect. There is full chaos, with
surjective mapping, for an infinite range of $B$. A property such
as this can be very useful in generating pseudo-random numbers. We
explore this application in Section \ref{sec:prng}.

Another interesting property of the B-exponential Map tends to a
constant function (with value 1) as B tends to $\infty$, for all
$x$.
\subsection{The Numerator Term}
\begin{figure}[t]
\centering
\includegraphics[scale=.6]{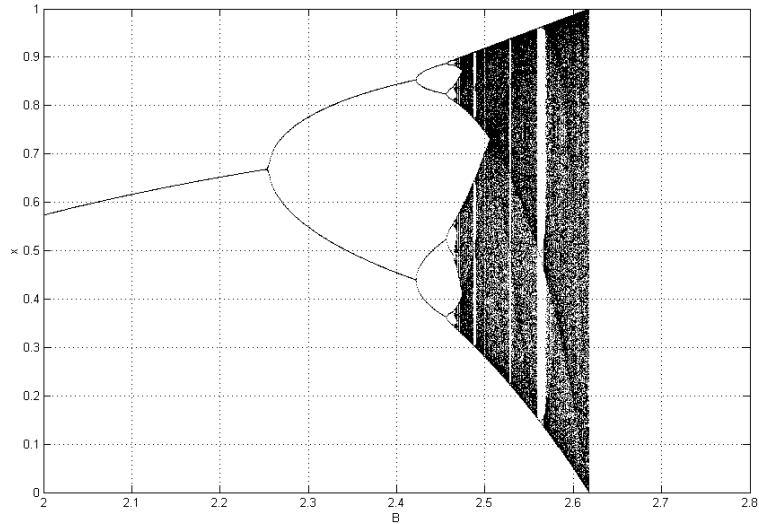}
\caption{A blown up version of the bifurcation diagram of the
numerator term $G(x)$. Full-chaos is achieved at
$B=\phi^2\approx2.6180$}. \label{fig:fig:bif num1}
\end{figure}

The numerator term of the B-exponential map,
$G(B,x)=B-xB^x-(1-x)B^{1-x}$ has interesting properties. It
becomes topologically conjugate to the logistic map at $B=\phi^2$
where $\phi$ is the golden mean ($\phi = \frac{1+\sqrt 5}{2}$).
Thus, it becomes fully chaotic at $B=\phi^2$. The bifurcation
diagram of the numerator term is shown in Figure \ref{fig:bif
num}. Figure~\ref{fig:fig:bif num1} shows the period doubling
route to chaos of the numerator term.
\begin{figure}[!hbp]
\centering
\includegraphics[scale=.6]{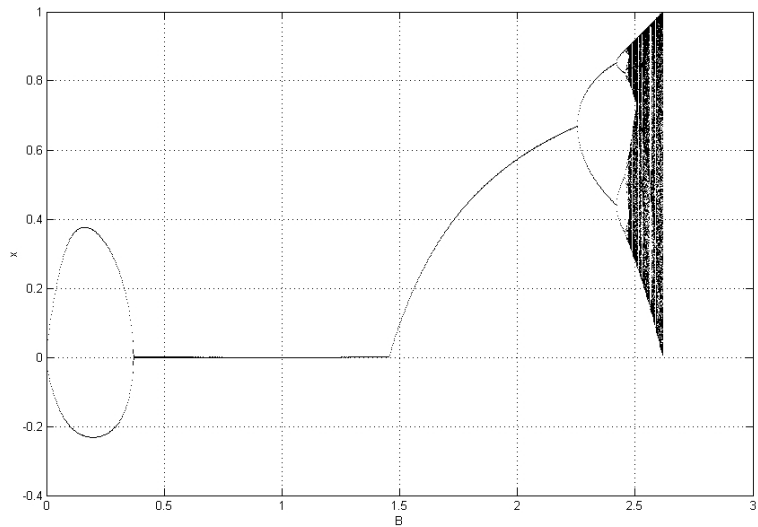}
\caption{Bifurcation diagram of the numerator term $G(x)$. The end
of chaos can be seen at $B=\phi^2\approx2.6180$.} \label{fig:bif
num}
\end{figure}

\section{A Pseudo-random Number Generator based on the B-Exponential Map}
\label{sec:prng} A random number is \emph{a number chosen as if by
chance from some specified distribution such that selection of a
large set of these numbers reproduces the underlying distribution}
\cite{mathworld}. An ideal random number generator is a discrete
memoryless information source that generates equiprobable symbols.
In this report, the phrase \emph{random numbers} refers to
uniformly distributed random numbers. Pseudo-random number
generators are algorithms implemented on digital systems that can
generate sequence of numbers which are random-like in their
statistical properties. Due to limitations in computation and
precision, pseudo-random number sequences are necessarily
periodic. Sequences generated by pseudo-random number generator
algorithms are expected to have large periods and pass a number of
statistical randomness tests.

Chaotic dynamical systems exhibit unpredictability, ergodicity and
mixing properties. This suggests that chaotic maps can be used in
generating random numbers. The relationship between chaos and
cryptography have been discussed by Kocarev~\cite{kocarev 1}.
Various one-dimensional chaotic maps have been proposed for
generating random numbers, e.g: PL1D~\cite{kocarev 2},
LOGMAP~\cite{logmap},etc. In their study of the Logistic Map as a
random number generator, Pathak and Rao~\cite{logmap} propose the
logistic map as a pseudo-random number generator which has a
period of about $10^8$ when implemented in double precision. This
period is quite small when compared to many other `good' random
number generators in the literature. They conjecture that such a
period is due to the fact that the value of $a$ in $y=ax(1-x)$
becomes slightly less than $4$ (which corresponds to full chaos),
because of which the map goes into periods. The fact that full
chaos exists only for a small set of parameters is a major
hindrance in using chaotic maps as pseudo-random number
generators. The limitations of computation and precision cause the
parameters to deviate from full chaos values and this result in
periodicity. Another major disadvantage of using chaotic maps
directly is that the the successive points are strongly
correlated. This shows up in the 2-dimensional phase space which
will be the plot of the mapping function.

\begin{figure}[!hbp]
\centering
\includegraphics[scale=.6]{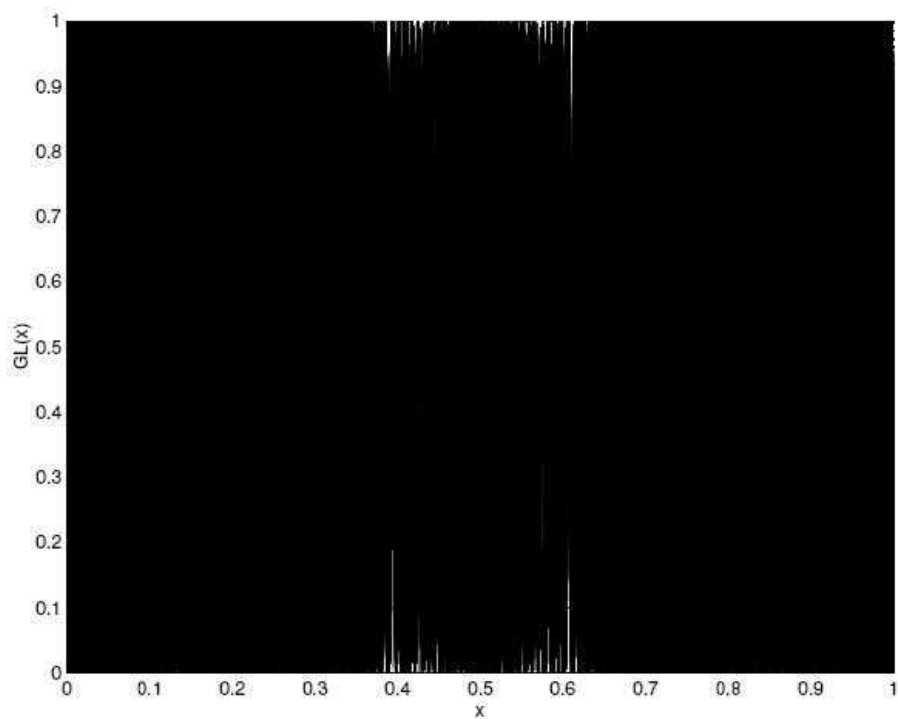}
\caption{The tenth return map for various values of $B \geq
e^{-4}$. } \label{fig:nth ret map}
\end{figure}

To overcome some of the problems mentioned above, one of the
strategies might be to take iterates from different maps and use
them as a sequence of random numbers. As shown in Theorem
\ref{thm:thm2}, the B-exponential Map shows full chaos for all
values of B greater than $e^{-4}$. This provides us with a
theoretically infinite number of maps to choose from. We
\emph{hop} from map to map, picking iterates on each hop to
generate random numbers. Figure~\ref{fig:nth ret map} shows the
10th return maps for a large number of B's $>e^{-4}$. It can be
seen that the maps almost fill the space even with just ten
iterates. This gives evidence that an algorithm based on map
\emph{hopping} on $B$ has the potential to yield a uniformly
distributed set of numbers between 0 and 1.

\subsection{BEACH}
We propose a PRNG based on B-Exponential Map with the name BEACH
(B-Exponential All-Chaotic map-Hopping). As the name suggests, the
pseudo-random number generator is based on the principle of
hopping from map to map to extract numbers for the generator. Such
a scheme has been studied by Rowlands~\cite{rowlands} and
Zhang~\cite{mmohocc}. Their methods were limited by the choice of
maps and the kind of hopping mechanism. MMOHOCC of
Zhang~\cite{mmohocc} uses a finite number of arbitrarily
predefined chaotic maps. They use pre-defined hopping patterns to
extract points from the trajectories. We propose a different
hopping mechanism, one that is deterministically chaotic. We also
have the advantage of choosing from a very large number of fully
developed chaotic maps. Zhang's MMOHOCC also has the problem of
not having {\it robust chaos} for any of their maps. The maps they
use (Chebyshev and Logistic) are not full-chaos for all values of
the parameter. They use the fully chaotic value of $a=4$ for the
Logistic Map. However, such a method would have the draw-back of
not being fully chaotic when implemented in hardware. It is
impossible to maintain a constant value for a parameter exactly in
hardware owing to noise. Another problem they have to worry about
is the presence of periodic orbits for some values of the
parameters even in the chaotic regime (this is indicated by the
windows in the bifurcation diagrams). Our method eliminates all
these problems.

We showed in Theorem 4 that B-exponential Map exhibits {\it robust
chaos} for all values of $B \geq e^{-4}$. In other words, for no
value of $B \geq e^{-4}$ does the B-exponential map settle into
periodic orbits (except for a measure zero set of initial
conditions). The BEACH uses this property to parametrically
generate many maps exhibiting full chaos. In the following
section, we discuss the BEACH algorithm.

\subsection{The Algorithm}
The seed to the BEACH pseudo-random number generating algorithm is
any number between 0 and 1. This number forms the initial value of
the iteration. Let $x_0$ denote the seed. We assume that the seed
itself is generated using a {\it random} procedure like the
movement of the mouse, the speed of typing on the keyboard or some
physical characteristic (like heat dissipation) in the hardware.
In BEACH, each random number is picked from a particular map. The
maps are generated parametrically using a sequence of $B$'s,
$\{B_1,B_2,....,B_M\}$ where $M$ is the number of maps we wish to
use for hopping. In our implementation, we pick one iterate from
each of the maps. Thus $M$ becomes equal to the length of the
pseudo-random number sequence we intend to generate (however the
$B$s are not necessarily distinct though there are a potentially
infinite number of them). Let $x_n$ denote the $n^{th}$ iterate of
the map corresponding to $B=B_m$. Since BEACH takes one value from
each iterate, the sequence of random numbers will be
$\{x_1,x_2,x_3,....,x_m\}$.
\\Let,
\begin{equation}
f(B_m,x_n)=
\frac{B_m-{x_n}B^{x_n}-(1-x_n){B_m}^{1-x_n}}{B_m-\sqrt{B_m}}\label{eq:beach}
\end{equation}
\\Where $B_m$ is the $m^{th}$ value of the
sequence $\{B_1,B_2,....,B_M\}$. This sequence of $B$'s can be
generated in many ways. We only need to ensure that successive
$B$'s are not sufficiently close with a high frequency so that any
two consecutive maps differ considerably. One way of varying $B$
is by using the Logistic Map. Alternatively, $B$ can also be
varied using a simple, weak PRNG like Linear Congruential
Generator. Such a scheme ensures that successive $B$'s are not
close to each other on the real line for most of the times. We
could also vary $B$ using the orbit of BEACH itself. We use the
Logistic Map for generating $B$'s, with slight modifications for
effective implementation. We explain the modifications in the
subsequent sections. Although varying $B$ according to the
logistic map does not give a random sequence of $B$'s, it is
sufficient for the purpose of hopping maps. The logistic map is
periodic because of limitations in precision. But this does not
bring about periodicity in the pseudo-random number sequence
because for each of the $B$'s, the previous iterate is different.

For a particular $B_m$, $f(B_m,x_{m-1})$ is iterated on $x_{m-1}$,
keeping $B_m$ constant. $R$ such iterations make sure that $x_{m}$
is considerably different from $x_{m-1}$, especially for $x_m$'s
close to the fixed points of the $B_m$ map. Now, the $R^{th}$
iterate is extracted as the new random number $x_m$. The same is
repeated for a new, updated value of $B$, $B_{m+1}$, i.e
$f(B_{m+1},x_{m})$ is iterated on $x_{m}$, $R$ times, with
$B_{m+1}$ remaining constant, and the $R^{th}$ iterate will be
$x_{m+1}$ and so on. $R$ is chosen such that there is a good
trade-off between time of computation and departure from previous
iterate. A value of $R=20$ should be sufficient to ensure good a
departure from $x_{m-1}$ even when $x_{m-1}$ is close to the fixed
point of $B_{m}$. The pseudo-code of BEACH is given below. We
limit the value of $B$ because for large values of $B$, as seen
from Figure \ref{fig:many maps}, the maps become flat in shape.
This may result in periodicity or fixed points owing to
limitations of precision on a computer (values very near to 1 may
be rounded off to 1 which becomes 0 in the next iterate).

\begin{tabbing}
~~~~~~~~~~~~~~~~~~~~~~~~$x_{old}=SEED$; \\
~~~~~~~~~~~~~~~~~~~~~~~~$y_{old}=SEED$;\\
~~~~~~~~~~~~~~~~~~~~~~~~for\_1(i\==1;$i<LEN/32$;i++) \\
\>    $y_{new}=4y_{old}(1-y_{old})$; \\
\>    $y_{old}=y_{new}$;          \\
\>   for\_2(j\==1;$j<R$;j++)\\
\>\>
$x_{new}=\frac{B-{x_{old}}B^{x_{old}}-(1-x_{old})B^{1-x_{old}} }
 {B-\sqrt{B}}$;\\
 \>\>     $x_{old}=x_{new}$;\\
 \>\>     $x_{newint}=x_{new}\times{2^{52}}$;\\
 \>   end\_for\_2\\

\>       if\_1($y_{new}\leq \frac{1}{BLIMIT}$)\\
\>\>        if\_2(\=$x_{new}\geq \frac{1}{BLIMIT}$)\\
\>\>\>        $y_{new}=x_{new}$;\\
\>\>        else\\
\>\>\>        $y_{new}=\frac{1}{BLIMIT}$;\\
\>\>  end\_if\_2\\
\>   end\_if\_1\\
~~~~~~~~~~~~~~~~~~~~~~~~  end\_for\_1\\
\end{tabbing}

\par The $SEED$ is the initial {\it random} seed input to the
program. In this particular implementation, $SEED$ cannot be
initialized to 0.75 since this is a fixed point of the Logistic
Map. The other disallowed seeds are 0 and 1 for obvious reasons.
$LEN$ is the number of bits which we wish to generate using the
PRNG. $BLIMIT$ refers to the maximum value of $B$ allowed.
\subsection{Period of BEACH}
It is very hard to analytically determine the period of BEACH.
Theoretically, robust chaos implies that there are no stable
periodic orbits and we also know that the measure of periodic
orbits is zero (in full chaos). However, when implemented on a
computer, all orbits are periodic owing to limited precision.
Since we have implemented BEACH in double precision arithmetic,
the number of chaotic maps available for hopping is around
$10^{300}$. As we are hopping in a chaotic fashion, consecutive
maps from which random numbers are extracted will be considerably
different. Given this, we believe that the period of BEACH will be
at least  $10^{300}$.
\section{Implementation}
The implementation of the algorithm was written in ANSI C in
double precision. Theoretically, a chaotic iteration may visit any
point in the interval $(0,1)$. This poses some problems in
practical implementations. If an iterate goes very close to zero,
less than the precision, it will be truncated to zero. We use a
\emph{zero-trap} to prevent such a thing from happening. The
\emph{zero-trap} traps values which are very close to zero and
replaces them with arbitrary iterates of the Logistic Map. These
are the same iterates which are used to generate $B$.

As stated earlier, $B$ is generated using the Logistic Map
recursion. The initial value for the recursion on $B$ is set equal
to the seed of the generator. The B-Exponential Map is quite
\emph{flat} for large values of $B$, as shown in Figure
\ref{fig:many maps}. It is desirable to limit the value of $B$ for
generating pseudo-random numbers. We choose the upper limit of $B$
as $10,000$. Hence, $B$ is extracted $10,000$ times from the
iterate of the Logistic Map with the seed as the initial value. If
an iterate is lesser than $10^{-4}$, it is replaced by an iterate
of the B-Exponential Map. If the iterate of B-Exponential Map is
less than $10^{-4}$, the Logistic Map iterate is set to $10^{-4}$.
Thus, we ensure that $B$ does not exceed $10,000$. The
B-exponential map iteration gives a value between 0 and 1, in
double precision in our implementation. To convert this to
integers, we multiply the iterate by $2^{52}$ (similar to Zhang's
method).

\begin{figure}[t]
\centering
\includegraphics[scale=.6,angle=270]{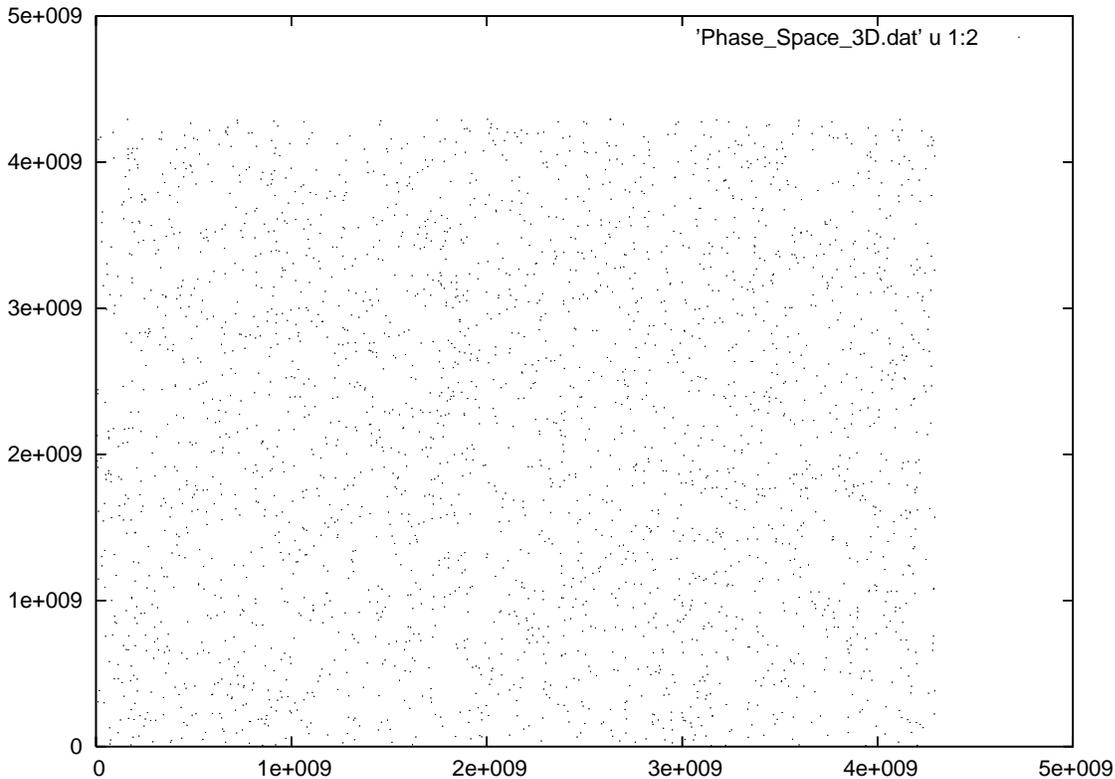}
\caption{The 2-dimensional phase space of a sequence of 3000
integers generated using BEACH.} \label{fig:phase space 2d}
\end{figure}

\begin{figure}[!hbp]
\centering
\includegraphics[scale=.6,angle=270]{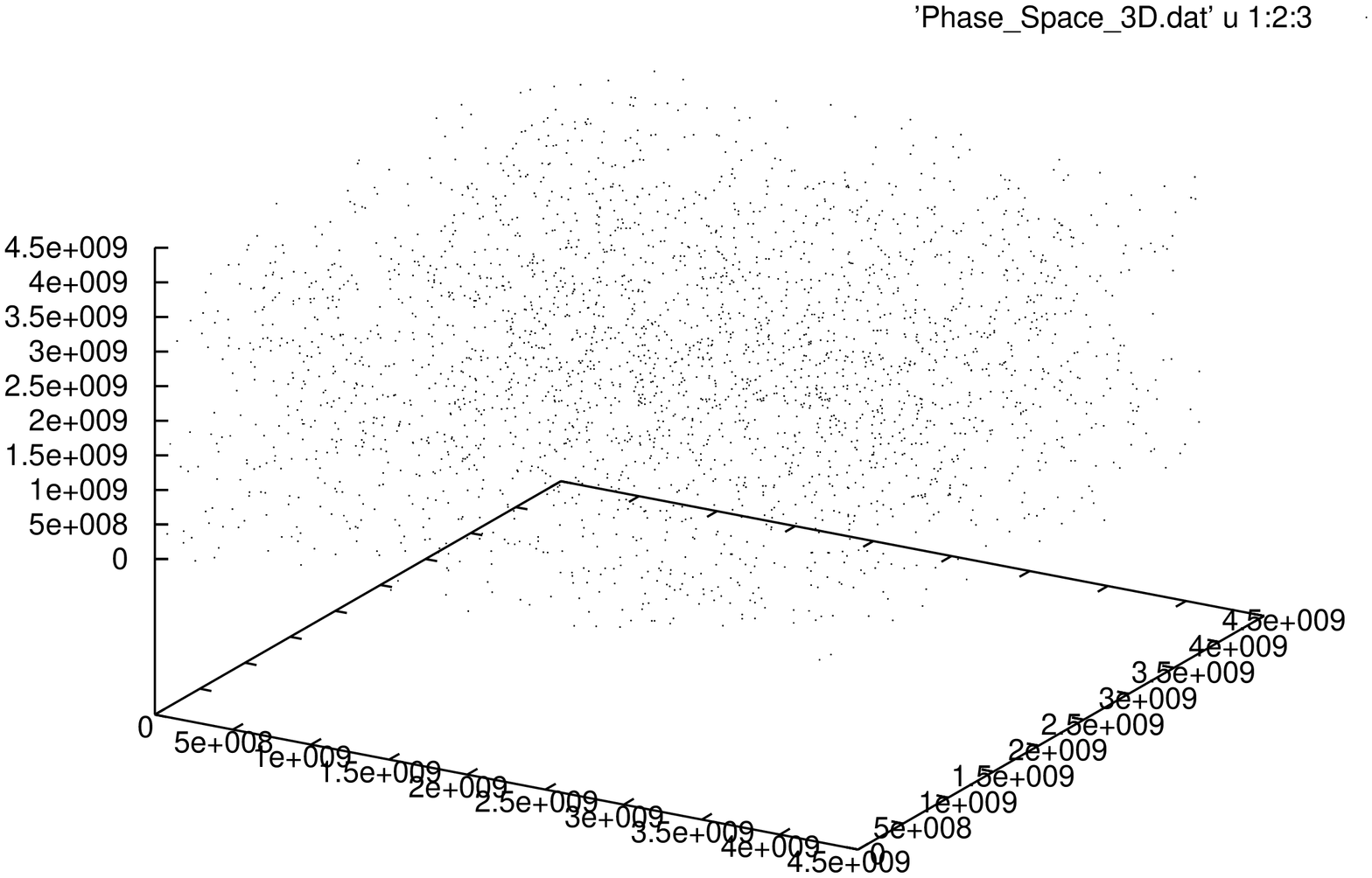}
\caption{The 3-dimensional phase space of a sequence of 3000
integers generated using BEACH.} \label{fig:phase space 3d}
\end{figure}

The 2-dimensional phase space of 3,000 random numbers generated
using BEACH is shown in Figure \ref{fig:phase space 2d} (i.e.
$x_{n}$ vs. $x_{n+1}$). The 3-dimensional phase space (i.e.
$x_{n}$ vs. $x_{n+1}$ vs. $x_{n+2}$) is shown in Figure
\ref{fig:phase space 3d}. It can be seen from the 2-dimensional
phase space that the numbers seem to randomly fill the entire area
between 0 and $2^{32}$ and the $2^{32}$ by $2^{32}$ by $2^{32}$
cube in the 3-dimensional phase space. This seems to provide a
strong evidence that the pseudo-random numbers are not correlated
in either 2 or 3-dimensions. Weak generators like the linear
congruential generator, distinctly settle in patterns in the
3-dimensional phase space.

\section{Randomness Evaluation of BEACH pseudo-random number generator}
The BEACH pseudo-random number generator was tested using 3
standard tests- The National Institute of Standards in
Technology's Statistical Test Suite (NIST)~\cite{nist, nist
manual}, George Marsaglia's Diehard Battery of
tests~\cite{diehard}, and the ENT test~\cite{ent test}. The BEACH
pseudo-random number generator successfully passed all the 15
tests of NIST and all the 18 tests of Diehard Battery.
\subsection{Entropy, Chi-square and Mean} Entropy (Shannon
entropy) is defined as $H(X)=-\sum_{x}P(x)\log_2{P(x)}$, where
$P(x)$ is the probability that the random variable $X$ is in the
state $x$. Thus, entropy is a measure of the \emph{information
density} of the data. We created a binary file with the random
numbers (taken as 32 bit integers). Thus, the entropy of such a
file would be defined based on the states 0 and 1. An optimal
compression of the file using the ENT Pseudo-random Number
Sequence Test Program (by John Walker)~\cite{ent test}, resulted
in an entropy of 1 per symbol. Thus, the program was unable to
compress the file. This gives strong evidence that BEACH is a good
pseudo-random number generator. This is supported by the fact that
the file also passed the Lempel-Ziv Compression test which is a
part of the NIST Statistical Testing Suite.

The chi-square test is a very basic test of randomness.
Knuth~\cite{knuth taocp 2} gives a detailed treatment of the
chi-square test. The chi-square distribution is for a sequence
file and expressed as an absolute number and a percentage which
indicates how frequently a truly random sequence would exceed the
value calculated. This percentage is a measure of the randomness.
If the percentage is less than 1\% or greater than 99\%, then the
sequence is not random. Percentages between 90\% and 95\% and 5\%
and 10\% indicate the sequence is ``almost suspect'' ~\cite{knuth
taocp 2}. Sequences generated by BEACH were within 25\% to 75\%
consistently.

The mean of 1 billion bit sequences was consistently at 0.5 for 1
bit word length and 127.5 for 8 bit word length. The serial
correlation was also very low, of the order of $10^{-5}$ for a
billion bit sequence. In addition to this, ENT program carried out
Monte Carlo Value of Pi test. Each successive sequence of 24 bits
are used as X and Y co-ordinates within a square. If the distance
of the randomly-generated point is less than the radius of the
circle inscribed within the square, the 24-bit sequence is
considered a \emph{hit}. The percentage of hits is used to
calculate the value of $\pi$. For very large streams (this
approximation converges very slowly), the value will approach the
correct value of $\pi$ if the sequence is close to random. For
BEACH, the error percentage was almost 0.00\%, consistently. For
the complete ENT test results, visit
$http://mahesh.shastry.googlepages.com/beach$.
The following table gives the value of these parameters for
different lengths of random bits. \\
\begin{table}
\begin{tabular}{|c|c|c|c|c|c|}
  \hline
  Length & Entropy & Chi-square  & Arithmetic & Monte Carlo & Serial\\
  &(per bit)&distribution(\%)& mean & value of $\pi$ (error \%)&correlation coeff
  \\
  \hline
  100 $Mb$ & 1.000000 & 50.00 & 0.5000 & 0.01 & 0.000151  \\
  500 $Mb$ & 1.000000 & 50.00 & 0.5000 & 0.00 & 0.000024 \\
  1 $Gb$   & 1.000000 & 75.00 & 0.5000 & 0.01 & 0.000035 \\
  \hline
\end{tabular}
\vspace{0.1in} \caption{Results of ENT on 3 bitstreams. }
\label{tab:tabent}
\end{table}
\subsection{NIST Statistical Test Suite} 
\begin{figure}[!hbp]
\centering
\includegraphics[scale=.6]{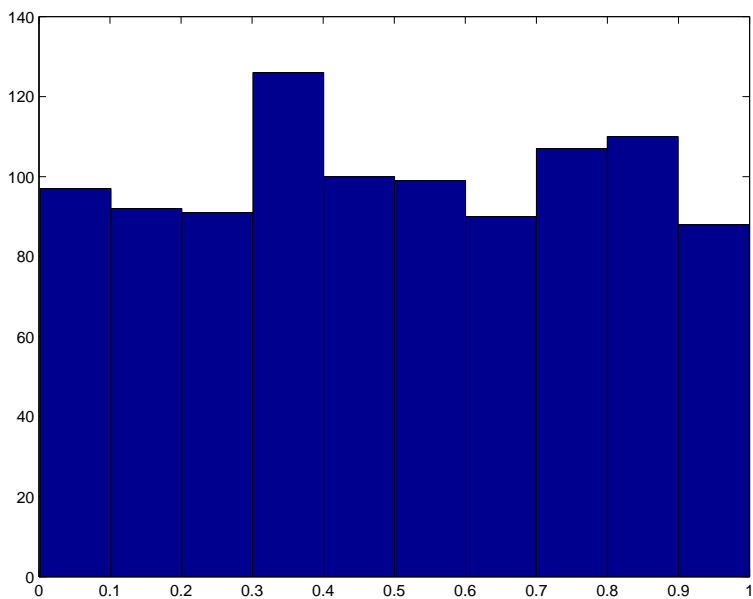}
\caption{The histogram of p-values of the block-frequency test of
NIST Suite. It indicates that the p-values are uniform.}
\label{fig:hist nist blok freq}
\end{figure}

BEACH random numbers also
passed the NIST Statistical Test Suite~\cite{nist}. The input was
given as an ASCII file consisting of 1's and 0's. The
pseudo-random number sequence passed all the 15 tests of the NIST
Suite. NIST checked the uniformity of p-values of 1000 streams of
1 million bits each and returns a p-value of the p-values. The
p-values of all the $15$ tests for each of the 1000 streams were
greater than 0.01 which is the limit for passing a test. It was
found that the p-values of p-values were all higher than 0.001.
This meant that the p-values are all uniform. \emph{Passing} of a
test in NIST Suite implies a confidence level of 99{\%}. In other
words, when the p-value is more than the passing level, the test
is considered passed with a confidence level of 99{\%}. The
details of all the 15 the tests and their interpretation is given
in~\cite{nist manual}. The histogram of p-values for the Block
Frequency Test are given in Figure~\ref{fig:hist nist blok freq}.
A plot of the proportion of bit-streams passing the NIST tests is
given in Figure~\ref{fig:nist threshold}, with the 2 thick lines
indicating the passing values.
\begin{figure}[!hbp]
\centering
\includegraphics[scale=.6]{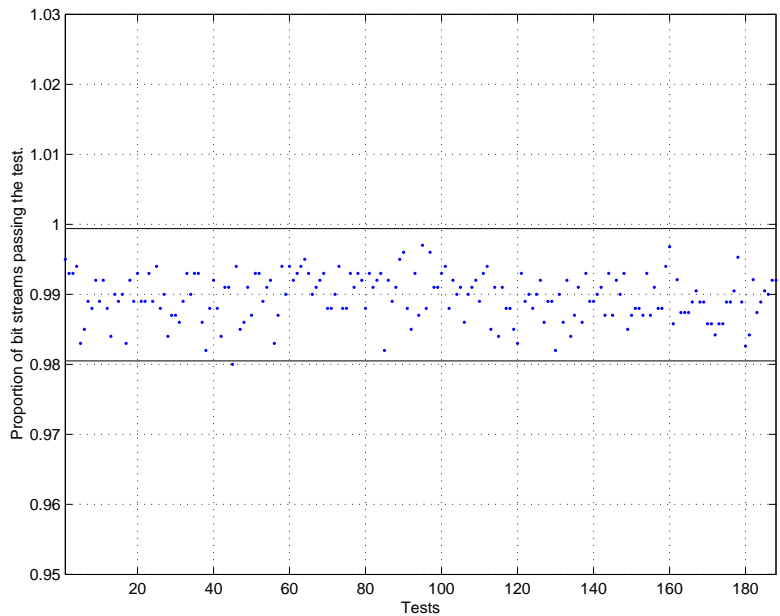}
\caption{The proportion of templates passing the test. The black
lines indicate the threshold for the passing value. The threshold
for passing is given approximately by $\alpha\pm
3\sqrt{\alpha(1-\alpha)/M}$ here, $\alpha=0.01$ is the passing
value. $M$ is the number of bitstreams.} \label{fig:nist
threshold}
\end{figure}
\subsection{Diehard Battery of Tests} The Diehard Battery of
\label{sec:diehard} Tests of George Marsaglia are collectively
considered to be one of the most stringent tests for randomness.
Ten streams of 1 billion bits each were generated using ten
different random seeds. Each of the seed was chosen randomly from
10 equally spaced intervals of (0,1). The criteria for passing a
Diehard test is that the p-value should not be 0 or 1 up to 6
decimal places. BEACH passed all the 18 tests recommended in the
Diehard Battery. The test results are tabulated in the table
below.
\\~~~
\begin{table}
\centering
\begin{tabular}{|c|c|c|c|c|c|c|c|c|c|c|c|}

  \hline
  Test & Seed & Seed  & Seed & Seed  & Seed  & Seed  & Seed  & Seed  & Seed  & Seed   \\
  No. & 1 & 2 & 3 & 4 & 5 & 6 & 7 & 8 & 9 & 10 \\
  \hline
  1 & .871  & .221 &  .250  & .922  & .014  & .154  & .069  & .050 & .790 & .515 \\
  2 & .971  & .527 &  .701  & .273  & .448  & .946  & .292  & .460 & .902 & .113 \\
  3 & .761  & .942 &  .321  & .679  & .407  & .585  & .519  & .587 & .448 & .962 \\
  4 & .374  & .801 &  .193  & .098  & .799  & .117  & .651  & .437 & .105 & .015 \\
  5 & .419  & .665 &  .317  & .324  & .040  & .876  & .678  & .507 & .468 & .075 \\
  6 & .925  & .199 &  .869  & .401  & .978  & .998  & .427  & .930 & .901 & .268 \\
  7 & .005  & .192 &  .935  & .611  & .505  & .621  & .729  & .339 & .125 & .773 \\
  8 & .516  & .549 &  .634  & .539  & .092  & .483  & .842  & .053 & .171 & .576 \\
  9 & .445  & .542 &  .080  & .964  & .724  & .773  & .807  & .136 & .383 & .806 \\
  10 &.244  & .084 &  .982  & .779  & .355  & .088  & .678  & .324 & .185 & .059 \\
  11 &.717  & .440 &  .045  & .269  & .280  & .376  & .542  & .873 & .589 & .952 \\
  12 &.566  & .702 &  .008  & .900  & .145  & .065  & .427  & .784 & .292 & .065 \\
  13 &.544  & .636 &  .096  & .376  & .768  & .922  & .324  & .903 & .987 & .677 \\
  14 &.804  & .835 &  .845  & .302  & .390  & .791  & .235  & .567 & .802 & .746 \\
  15 &.082  & .512 &  .523  & .875  & .713  & .604  & .704  & .909 & .482 & .119 \\
  \hline
\end{tabular}
\vspace{0.1in}
 \caption{Results of Diehard battery of tests on 10 bitstreams of 1 Gb each. The names of the tests are given in Appendix B.} \label{tab:tabdiehard}
\end{table}
The full results, for all the seeds, of the Diehard Tests are
available at
\\
$http://mahesh.shastry.googlepages.com/beach$.
\newline{A very few of the results were close to 0.9999. For those
seeds, longer sequences were tested. The results of the Diehard
Tests for longer sequences showed that such p-values close to
1.00000 were one-off occurrences. We have found that BEACH passes
all the tests of Diehard and NIST for extremely large sequences
(up to 10 Gb).}
\section{Summary and Future Research Directions}
We have for the first time explored the utility of {\it robust
chaos} for the generation of pseudo-random numbers. By chaotically
hopping across maps which are fully chaotic, BEACH appears to be a
strong candidate for pseudo-random number generators for
cryptographic purposes with very long periods. The fact that BEACH
successfully passes stringent statistical tests even for extremely
long sequences is a testimony to this statement.
The existence of one of the following, a secure pseudo-random
number generator, a secure block encryption algorithm, and a
secure one-way function, implies that the other two also
exist~\cite{kocarev 1}. This makes a strong case for the search of
strong pseudo-random number generators and we believe that {\it
robust chaos} will have an important role to play. Our future
research would focus on developing a strong stream cipher based on
BEACH. To this end, we wish to perform security analysis on BEACH.
\section*{Acknowledgements}
We would like to thank Sutirth Dey of the Jawaharlal Nehru Centre
for Advanced Scientific Research for stimulating discussions on
the B-Exponential Map. Nithin Nagaraj would like to express his
sincere gratitude to the Department of Science and Technology
Ph.D. fellowship program and to Timothy Poston, National Institute
of Advanced Studies for discussions on topological conjugacy.

\newpage
\section*{Appendix}
\appendix
\section{Derivation of certain results}
\subsection{Derivation of generalization of the
Logistic Map} To prove that
\begin{equation}
 \lim_{B \rightarrow 1} GL(B,x)=4x(1-x).
\end{equation}
i.e.;
\begin{equation}
\qquad \lim_{B \rightarrow 1} \frac{B-xB^x-(1-x)B^{1-x} }
{B-\sqrt{B}}=4x(1-x).
\end{equation}
\\ Consider the LHS; by applying L'Hospital's rule;
\begin{equation}
\lim_{B \rightarrow 1} \frac{B-xB^x-(1-x)B^{1-x} }
{B-\sqrt{B}}=\lim_{B \rightarrow 1}
\frac{1-x^2B^{x-1}-(1-x)^2B^{-x} } {1-1/2\sqrt{B}}=4x(1-x).
\end{equation}
\subsection{B-exponential map loses surjectivity at $1/e^4$}
For surjectivity, the local maximum of the map in the
interval $x=(0,1)$ should be equal to 1. Using this condition, we
show that the map loses surjectivity for $B<e^{-4}$. The condition
for extremum of $GL(x)$ with respect to $x$ is:
\begin{equation}
\frac{\partial{GL(B,x)}}{\partial{x}}=0.
\end{equation}
Now, we know that the B-exponential map is symmetric and concave.
Hence, $x=0.5$ should be a maximum or a minimum point. Thus, we
get:
\begin{equation}
\frac{\partial{GL(B,x)}}{\partial{x}} \Big|
 _{x=0.5}=0\qquad
\forall~B \in \mathbb{R^+}.
\end{equation}
i.e:
\begin{equation}
\frac{\partial{ \Big(\frac{B-xB^x-(1-x)B^{1-x} } {B-\sqrt{B}}
 \Big) }}{\partial{x}} \Big| _{x=0.5}=0\qquad \forall ~ B \in
\mathbb{R^+}.
\end{equation}
But, when $0.5$ is a local minimum in any interval, then
$GL(B,0.5)=1$ will be a local minimum and the map will no longer
be surjective. We use this condition to find $B$ for which the map
loses surjectivity.
\begin{equation}
\frac{\partial^2{GL(B,x)}}{\partial{x}^2} \Big| _{x=0.5}>0.
\end{equation}
Applying this, we get for $B<e^{-4}$, B-exponential map is not
surjective.

This also shows that the second derivative of the function with
respect $x$, is negative for all values of $x$ only when
$0<B<e^{-4}$.
\subsection{The B-exponential Map becomes constant at 1 as B tends to $\infty$}
To find;
\begin{equation}
\lim_{B \rightarrow \infty} \frac{B-xB^x-(1-x)B^{1-x} }
{B-\sqrt{B}}.
\end{equation}
rearranging the terms, we get;
\begin{equation}
\lim_{B \rightarrow \infty} \frac{1-xB^{(x-1)}-(1-x)B^{-x} }
{1-1/\sqrt{B}}=1.
\end{equation}
Hence proved.
%
%
%\subsection{The B-exponential Map is bounded by the Logistic Map from below}
%We shall prove here that the B-exponential Map is bounded by the
%Logistic Map from below. This means that for every $B > e^{-4}$,
%the curve of $GL(B,x)$ is above the Logistic Map.
%
%
%
\subsection{Generalized Tent Map}
We know that the Logistic Map is conjugate to the Tent
Map~\cite{yorke}. We can use the explicit conjugacy map on
$GL(B,x)$ to obtain a generalization of the Tent Map. We know that
the conjugate map $C(x)$ is given by~\cite{yorke}:
\begin{equation}
C(x) = \frac{(1-cos~\pi x)}{2} = sin^2 {\Big(} \frac{\pi x}{2}
{\Big)}.
\end{equation}
The Generalized Tent Map is given by:
\begin{equation}
GT(B,x) = \frac{2}{\pi} sin^{-1} {\Big(} \frac{ B -
sin^2(\frac{\pi x}{2})B^{sin^2(\frac{\pi x}{2})}  -
(1-sin^2(\frac{\pi x}{2}))B^{(1-sin^2(\frac{\pi x}{2}))} }{B -
\sqrt{B}} {\Big)^{\frac{1}{2}}}.
\end{equation}
with the following property:
\begin{equation}
\lim_{B \rightarrow 1} GT(B,x) = 1-2 {\Big|} x-\frac{1}{2} {\Big|}
.
\end{equation}
The Generalized Tent Map is plotted below for a few values of B.
\begin{figure}[!hbp]
\centering
\includegraphics[scale=.4]{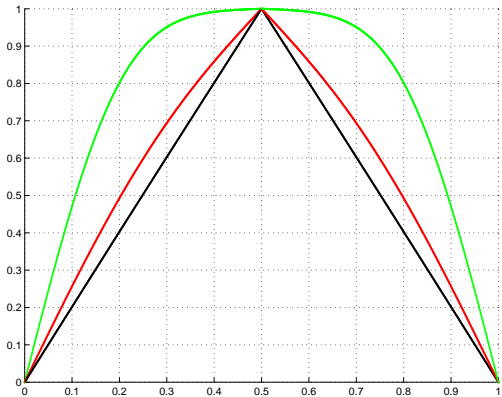}
\caption{The Generalized Tent Map for three values of $B$. Black:
$B=2$, Red: $B=200$, Green: $B=2 \times 10^{10}$.}
\label{fig:gtent}
\end{figure}
\par The Generalized Tent Map also exhibits {\it robust chaos}
property (since it is topologically conjugate to the B-Exponential
Map) and could also be used for generating PRNG.
\newpage
\section{List of tests included in NIST and Diehard suites}
\subsection{NIST tests:}
\begin{enumerate}
    \item   Frequency
    \item   Block-Frequency
    \item   Cumulative-Sums
    \item   Runs
    \item   Longest-Run
    \item   Rank
    \item   FFT
    \item   Nonperiodic-Templates
    \item   Overlapping-Templates
    \item   Universal
    \item   Approximate Entropy
    \item   Random-Excursions
    \item   Random-Excursions-Variant
    \item   Serial
    \item   Linear-Complexity
\end{enumerate}
\subsection{Diehard tests:}
\begin{enumerate}
    \item   Birthday Spacings Test
    \item   Overlapping 5-Permutation Test
    \item   Binary Rank Test For $31 \times 31$ Matrices and $32 \times 32$ Matrices
    \item   Binary Rank Test For $6 \times 8$ Matrices
    \item   Bitstream Test
    \item   Tests OPSO, OQSO And DNA
    \item   Count-The-1'S Test On A Stream Of Bytes
    \item   Count-The-1'S Test For Specific Bytes
    \item   Parking Lot Test
    \item   Minimum Distance Test
    \item   3Dspheres Test
    \item   Squeeze Test
    \item   Overlapping Sums Test
    \item   Runs Test
    \item   Craps Test
\end{enumerate}


\begin{thebibliography}{1}
\bibitem{yorke}
K T Alligood, J A Yorke, T D Sauer: Chaos: An Introduction to
Dynamical Systems; Springer-Verlag Inc., 1997
\bibitem{Barreto}
E Barreto, B R Hunt, C Grebogi, J A Yorke: From High Dimensional
Chaos to Stable Periodic Orbits: The Structure of Parameter Space,
Physics Review Letters, VOL. 78, NO. 24, June 1997
\bibitem{Banerjee}
S Banerjee, J A Yorke, C Grebogi: Robust Chaos, Physics Review
Letters, VOL. 80, NO. 14, April 1998
\bibitem{Andrecut}
M Andrecut, M K Ali: Robust chaos in smooth unimodal maps,
Physical Review E, VOL. 64, No. 025203, July 2001
\bibitem{Andrecut2}
M Andrecut, M K Ali: Example of robust chaos in a smooth map,
Europhyics Letters, VOL. 54 NO. 3, May 2001
\bibitem{knuth taocp 2}
D Knuth: The Art of Computer Programming, Volume 2: Seminumerical
Algorithms, Third Edition; (Reading, Massachusetts:
Addison-Wesley, 1997), xiv+762pp. ISBN 0-201-89684-2
\bibitem{kocarev 1}
L Kocarev: Chaos Based Cryptography: A Brief Overview, Circuits
and Systems Magazine; IEEE, 2001
\bibitem{logmap}
S C Pathak, S Suresh Rao: Logistic Map, A possible random number
generator; Physical Review E, VOL. 51, NO. 4, April 1995
\bibitem{kocarev 2}
T Stojanovski, L Kocarev: Chaos-Based Random Number
Generators—Part I: Analysis; IEEE Transactions on Circuits and
Systems-I: Fundamental Theory and Applications, VOL. 48, NO. 3,
March 2001
\bibitem{kocarev 3}
L Kocarev, G Jakimoski: Pseudorandom Bits Generated By Chaotic
Maps; IEEE Transactions on Circuits and Systems-I: Fundamental
Theory and Applications , VOL. 50, NO. 1, January 2003
\bibitem{finland rngs}
I Vattulainen, K Kankaala, J Saarinen, and T Ala-Nissila: A
Comparitive Study of Some Pseudorandom Number Generators;
arXiv:hep-lat/9304008 v2 10 Aug 1993
\bibitem{rowlands}
T Rowlands, D Rowlands: A More Resilient Approach to Chaotic
Encryption, ICITA2002
\bibitem{mmohocc}
X Zhang, L Shu, K Tang: Multi-Map Orbit Hopping Chaotic Stream
Cipher; http://www.arxiv.org/ftp/cs/papers/0601/0601010.pdf
\bibitem{mathworld}
W E Weisstein: ``Random Number'', From MathWorld--A Wolfram Web
Resource, http://mathworld.wolfram.com/RandomNumber.html
\bibitem{nist}
National Institute of Standards and Technology, Random Number
Generation and Testing; http://csrc.nist.gov/rng
\bibitem{nist manual}
A Rukhin {\it et. al.}: A Statistical Test Suite for Random and
Pseudorandom Number Generators for Cryptographic Applications;
NIST Special Publication 800-22, with revisions dated May 15, 2001
\bibitem{ent test}
John Walker: ENT: A Pseudorandom Number Sequence Test Program;
http://www.fourmilab.ch/random/
\bibitem{diehard}
G Marsaglia: The Diehard Battery of Tests of Randomness, available
online at: http://www.csis.hku.hk/~diehard/index.html
%
%
\end{thebibliography}
\end{document}